\documentclass[prl,twocolumn,showpacs,amsmath,amssymb,superscriptaddress]{revtex4}
\usepackage{graphicx}
\usepackage{dcolumn}
\usepackage{bm}
\usepackage{latexsym}
\begin{document}
%\preprint{APS/123-QED}

\title{Continuous Weak Measurement and Nonlinear Dynamics in a Cold Spin Ensemble}
\author{Greg A. Smith}\affiliation{Optical Sciences Center, University of Arizona, Tucson, AZ 85721}
\author{Souma Chaudhury}\affiliation{Optical Sciences Center, University of Arizona, Tucson, AZ 85721}
\author{Andrew Silberfarb}\affiliation{Department of Physics and Astronomy, University of New Mexico, Albuquerque, NM 87131}
\author{Ivan H. Deutsch}\affiliation{Department of Physics and Astronomy, University of New Mexico, Albuquerque, NM 87131}
\author{Poul S. Jessen}\affiliation{Optical Sciences Center, University of Arizona, Tucson, AZ 85721}
\date{\today}

\begin{abstract}
A weak continuous quantum measurement of an atomic spin ensemble can be implemented via Faraday rotation of an off-resonance probe beam, and may be used to create and probe nonclassical spin states and dynamics. We show that the probe light shift leads to nonlinearity in the spin dynamics and limits the useful Faraday measurement window.  Removing the nonlinearity allows a non-perturbing measurement on the much longer timescale set by decoherence.  The nonlinear spin Hamiltonian is of interest for studies of quantum chaos and real-time quantum state estimation.
\end{abstract}

\pacs{42.50.Ct, 32.80.-t, 03.65.Ta}   % PACS,

\maketitle

The process of quantum measurement involves a fundamental tradeoff between information gain and disturbance. In
a projective measurement, this backaction is strong enough to
collapse the state of the system and disrupt its coherent
evolution. In more realistic scenarios, the system is weakly
coupled to a probe, which is then measured to gain small amounts
of information at the cost of modest perturbation. Continuous
versions of this weak measurement scheme are of particular
interest in the context of real-time feedback control and the
creation and probing of non-classical states and dynamics \cite{one}.
Generally, the coupling of a probe to a single quantum system is so weak that the signal carrying information about the system becomes masked by probe noise. The signal-to-noise ratio of the measurement can be improved by coupling the probe to an ensemble of identically prepared systems, while at the same time the backaction on individual ensemble members can be kept low.  Of course the many-body system is now described by a collective quantum state, and when the measurement strength is sufficient to resolve the quantum fluctuations associated with a collective observable, backaction will be induced on the collective state and the uncertainty of the measured value can be squeezed \cite{two}.  The creation of such quantum correlation has applications in precision measurement and quantum information processing. \cite{three}

In this letter we use the linear Faraday effect to
probe the spins in an ensemble of laser cooled Cs atoms.\cite{four,five,six} Our
setup employs a probe beam tuned near the $D_{2}$ transition at
852 nm, whose linear polarization is rotated by an angle
proportional to the net spin component along the propagation axis.
Measuring the rotation with a shot-noise limited polarimeter
provides a weak measurement of the ensemble averaged spin in real
time. If the sample is optically thick
on resonance, the atom-probe coupling becomes strong enough to
allow the collective spin to be measured with resolution below the
quantum uncertainty of a many-body spin-coherent state, making it
possible to generate quantum correlations within the ensemble.
In the limit of large probe detuning, Faraday rotation has been
employed as a quantum non-demolition (QND) measurement of the
collective spin, and much interest has been focused on its ability
to generate spin squeezed states \cite{seven,eight}, to perform sub-shot noise
magnetometry \cite{nine} and to entangle separated spin ensembles. \cite{ten} We show that the dynamics of individual atomic spins is dominated by a nonlinear term in the light shift generated by the Faraday probe, which leads to rapid dynamical collapse and revival of the mean spin during Larmor precession. The collapse is generally much faster than decay due to optical pumping, and therefore determines the useful measurement window. Furthermore, both the nonlinearity and measurement strength are proportional to the rate of probe photon scattering \cite{fifteen,six}, making the integrated backaction independent of probe intensity and detuning. The nonlinearity will therefore play the dominant role in limiting backaction-induced effects such as spin squeezing and entanglement. We demonstrate that removal of the nonlinearity allows the non-perturbing nature of the Faraday measurement to be recovered on the longer timescale set by photon scattering.  In our example we measure Larmor precession and achieve an effective cancellation by orienting the probe polarization at a specific angle with the applied magnetic field.  This simple approach may be broadly useful in experiments that involve laser probed or trapped spin ensembles in uniform magnetic fields.

\begin{figure}[t]\resizebox{8.75cm}{!}{\includegraphics {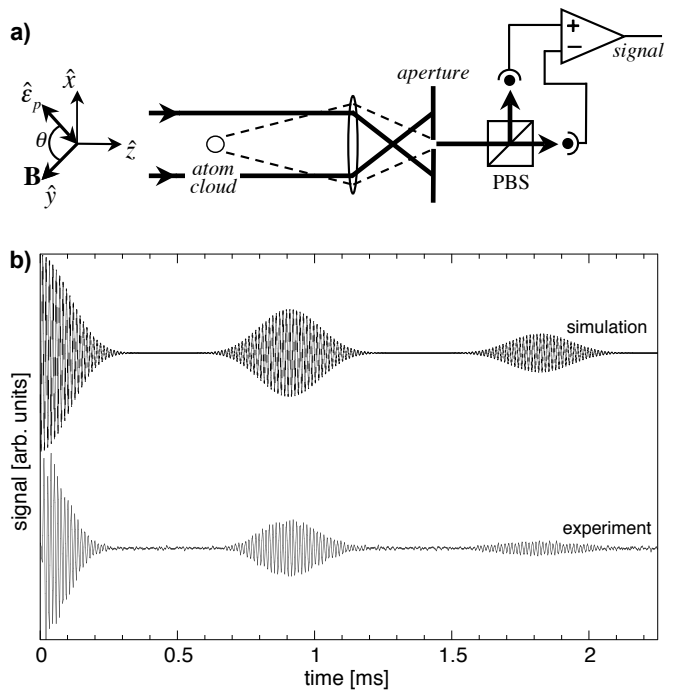}}
\caption{\label{fig:setup}(a) Experimental setup for Faraday rotation measurements in a sample of laser cooled Cs atoms.  (b) Master equation calculation and experimental data for an angle $\theta=0^{\circ}$ between the probe polarization and Larmor field, both showing collapse and revivals of the Faraday signal due to the nonlinear term in the single-spin Hamiltonian.}
\end{figure}

A general discussion of Faraday measurements in samples of laser
cooled alkali atoms can be found in \cite{six}, including sensitivity versus photon scattering tradeoffs and
requirements for significant backaction. In the following we consider corrections to the usual assumption of a non-perturbing measurement that arise due to the probe-induced light shift at large detuning. The light shift depends on the probe electric field and the atomic tensor polarizability, $\hat{U}=-\mathbf{E}_{p}^{(-)} \cdot\tensor{\alpha}\cdot \mathbf{E}_{p}^{(+)}$, and generally consists of scalar, vector and rank-2 tensor components. For detunings much larger than the excited state hyperfine splitting, $\Delta \gg \Delta _{HF}$, the scalar and vector components scale as $1/\Delta $, the vector component is identically zero when the probe polarization $\vec{\varepsilon} _{p}$ is linear, and the rank-2 tensor component scales as $1/\Delta ^{2}$\cite{eleven}. Keeping terms to leading order in $\Delta _{HF}/\Delta $
then yields a light shift\begin{equation}\label{eqn:lightshift}
\hat{U}=\frac{2}{3}U_{0}\hat{I}+\frac{\beta \ U_{0}}{\Delta
/\Gamma }\left( \vec{\varepsilon}_{p}\cdot\mathbf{\hat{F}}\right) ^{2}
\end{equation}where $\mathbf{\hat{F}}$ is the total angular momentum, $\beta $ is a numerical constant depending on the atomic
species, and where $U_{0}=s\hbar \Delta /2$\ is the light shift of
a two-level atom with unit oscillator strength, natural linewidth
$\Gamma $\ and saturation parameter $s=\left( \Gamma /2\Delta
\right) ^{2}\left( I_{p}/I_{0}\right) $\ for probe
intensity $I_{p}$. Introducing the probe photon scattering rate
$\gamma _{s}=s\Gamma /2$, and substituting the relevant parameters
for Cs in the $F=4$ hyperfine ground state, we find $\beta
U_{0}/\left( \Delta /\Gamma \right) \approx -1.2\gamma _{s}/\hbar
$. Finally, in the presence of a magnetic field \textbf{B},
we obtain a single-spin Hamiltonian \begin{equation}\label{eqn:hamiltonian}
\hat{H}=g_{F}\ \mu _{B}\ \mathbf{B}\cdot\mathbf{\hat{F}
}-1.2\left( \gamma _{s}/\hbar \right) \left( \vec{\varepsilon}_{p}\cdot\mathbf{\hat{F}}\right) ^{2}
\end{equation}where we have omitted the scalar (spin-independent) part of the
light shift. In addition to the Larmor interaction we
see here a nonlinear term that gives rise to dynamics
beyond simple rotations and leads to the generation of
non-classical spin states. This interaction has been studied in
the context of the ``kicked top'', a standard paradigm for quantum chaos \cite{twelve}, and leads to phenomena such as alignment-to-orientation conversion in polarization spectroscopy \cite{thirteen}.  More generally, it will occur in a variety of laser traps where its effects on the evolution of the atomic spins should be considered. In our
case the nonlinear level splitting induces rapid collapse and
subsequent revivals of the mean spin of a Larmor precessing spin
coherent state. We model this behavior in detail by setting up and
numerically solving a master equation for the Cs hyperfine ground
manifold, thereby fully accounting for both the coherent spin
dynamics of eq.\ (\ref{eqn:hamiltonian}) and for decoherence from optical
pumping. Figure\ \ref{fig:setup}b is an example of the calculated
expectation value $\langle  \hat{F}_{z}\rangle$ as a function of time, clearly showing both an initial Gaussian envelope
collapse and multiple revivals whose amplitudes are
limited by decoherence.

The spin dynamics depend not only on the magnitude of the
nonlinear
term, but also on the relative orientation of the applied field \textbf{B} and the probe polarization $\vec{\varepsilon}_{p}$. Consider the
geometry shown in fig.\ \ref{fig:setup}a, where \textbf{B} is orthogonal to the probe propagation and forms an angle $\theta
$ with $\vec{\varepsilon}_{p}$. Choosing the $y$-axis along
\textbf{B} we have $\left( \vec{\varepsilon}_{p}\cdot\mathbf{\hat{F}}\right) ^{2}=\left( \sin
\theta \  \hat{F}_{x}+\cos \theta \  \hat{F}_{y}\right) ^{2}$. For magnetic
fields $B\gg \hbar \ \gamma _{s}/g_{F}\ \mu _{B}$\ it is
appropriate to make a rotating wave approximation, and we obtain an effective nonlinearity\begin{equation}\label{eqn:nonlinear}
\left( \vec{\varepsilon}_{p}\cdot\mathbf{\hat{F}}
\right) _{RWA}^{2}=\left( -\frac{1}{2}\sin ^{2}\theta +\cos ^{2}\theta
\right)  \hat{F}_{y}^{2}
\end{equation}From this we see that the collapse and revival
occurs twice as fast at $\theta =0^{\circ}$ compared to $\theta
=90^{\circ}$, and that at a critical angle, $\theta =\arctan \left(\sqrt{2}
\right) \approx 54^{\circ}$, the nonlinearity disappears entirely. The
corresponding changes in dynamical behavior are clearly visible in our master equation
calculations.

Our experimental setup for Faraday measurements is similar to that
described in detail in \cite{six}. We begin by preparing a sample of a few
million Cs atoms in a magneto-optic trap,
followed by laser cooling in a 3D optical molasses and a 1D near-resonance optical lattice aligned along the
probe direction. Finally, the atoms are optically pumped to
produce a spin-coherent state within the $F=4$ ground manifold. Our probe beam is generated by a tunable diode
laser, spatially filtered by a single-mode optical fiber and
passed through a high quality Glan-Laser polarizer before it is
used to probe the atomic sample. The probe intensity profile is
very close to Gaussian with a $1/e$ radius of $\sim$ 1.2 mm.
This is significantly larger than the typical 0.25 mm radius of
the atomic cloud, and ensures that the probe light shift is
reasonably uniform across the ensemble. We use an imaging system
to select only the part of the probe that passes through the
cloud, and analyze it with a simple 
polarimeter consisting of a polarization beamsplitter and a
differential photo detector (fig.\ \ref{fig:setup}a). The resulting measurement
of the collective spin has a typical sensitivity that falls short
of the requirement for backaction by a factor $\sim 100$,
according to the estimate in \cite{six}. This puts us safely in the regime
of ensemble-averaged single-spin dynamics and backaction-free
measurements.

To observe the nonlinear spin dynamics of eq.\ (\ref{eqn:hamiltonian}) we rotate the
atomic spin coherent state to point along the $x$-axis, and apply the
Larmor field along the $y$-axis. These coordinate axes are chosen so
\textbf{B} forms the desired angle $\theta $ with the
(space fixed) probe polarization. We then measure the
time-dependent Faraday rotation, which gives a measure of
the ensemble average $\langle \hat{F}_{z}\rangle$. To
improve our signal-to-noise-ratio this process is repeated and the
measurement averaged 128 times. A typical result is shown
in fig.\ \ref{fig:setup}b, clearly demonstrating the rapid collapse and
revival of the Faraday signal in good agreement with our
theoretical model. Of particular note is the approximately
Gaussian envelope, which is qualitatively different from the
exponential envelope seen when the signal decays due to
optical pumping. The
experimental revival amplitude is typically
$\sim65\%$ of that predicted by theory, suggesting an extra source of photon scattering from e.g.\ amplified
spontaneous emission in the probe laser, or possibly small deviations
from the ideal linear probe polarization.

We have carried out measurements of the type in fig.\ \ref{fig:setup}b for a wide
range of probe intensity and detuning. \begin{figure}[t]
\resizebox{8.75cm}{!}{\includegraphics{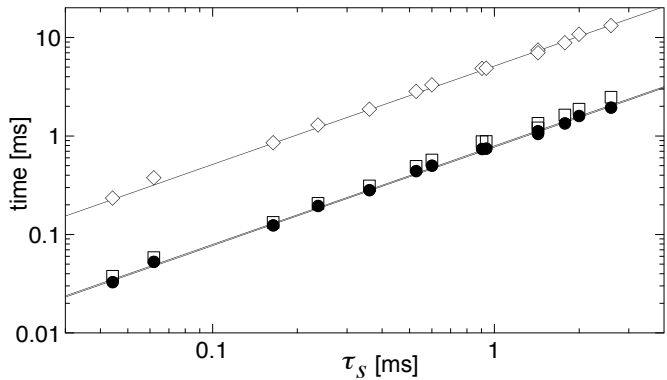}}
\caption{\label{fig:timings}Characteristic times for nonlinear collapse and revival in the Faraday signal, measured for $\theta=90^{\circ}$ and as a function of the scattering time $\tau_{s}=\gamma_{s}^{-1}$. ($\bullet$): time required for collapse to $1/e$ of the initial signal amplitude, ($\Diamond$) time of maximum revival amplitude, ($\Box$) time required for the first revival to collapse to $1/e$ of its maximum amplitude. Solid lines show the values predicted by a master equation calculation.}
\end{figure} Figure\ \ref{fig:timings} shows the observed
collapse and revival times obtained by fitting the signal with a Gaussian envelope, versus the photon scattering time $\tau _{s}=\gamma
_{s}^{-1}$. Our data is in good
agreement with theory and confirms the expected scaling behavior
over two orders of magnitude in $\tau _{s}$. Note that
$\tau _{s}$\ for an experimental data point must be calculated
from the probe intensity at the atomic sample and therefore is
somewhat uncertain. Here and elsewhere in the paper we scale the
experimental scattering times within a data set by a common factor to obtain the
best fit with theory; in general the inferred values for $\tau_{s}$ agree to
better than $20\%$ with those calculated from an independent
estimate of the probe intensity.

\begin{figure}[t]
\resizebox{8.75cm}{!}{\includegraphics{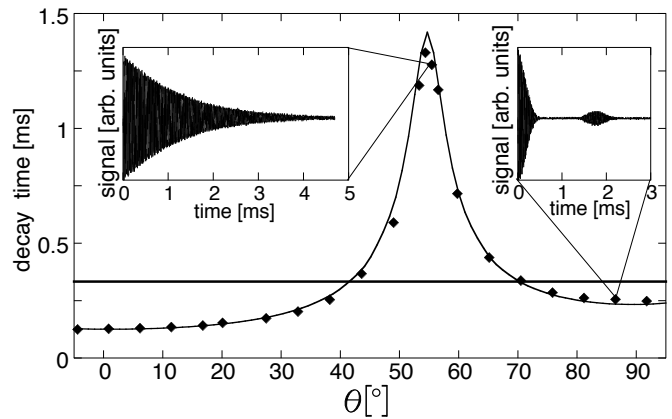}}
\caption{\label{fig:angles}$1/e$ decay times for the Faraday signal as a function of the angle $\theta$ between the probe polarization and Larmor field.  The solid curve shows the $1/e$ decay times predicted by a master equation calculation, and the horizontal line indicates the fixed scattering time used for this data set.  Inserts show the nonlinear collapse and revival at $\theta\sim90^{\circ}$, and the much longer exponential decay seen at the critical angle $\theta\approx54^{\circ}$.}
\end{figure}

\begin{figure}[t]
\resizebox{8.75cm}{!}{\includegraphics{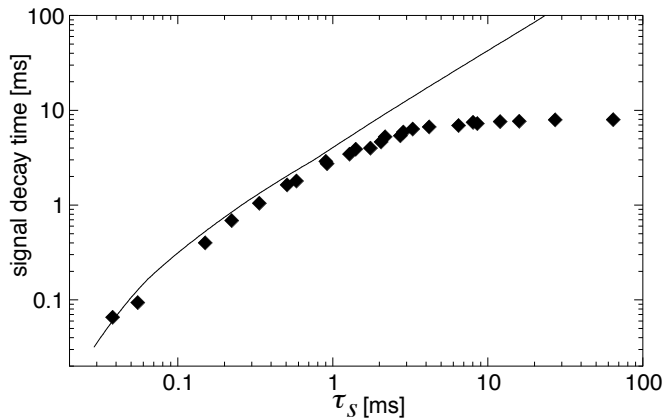}}
\caption{\label{fig:critical}$1/e$ decay times measured at the critical angle $\theta\approx54^{\circ}$, compared to the prediction of a master equation calculation (solid line). The experimental values reach a plateau at large $\tau_{s}$ due to ensemble inhomogeneity. The theory prediction deviates slightly from a straight line due to a partial breakdown of the rotating wave approximation.}
\end{figure}

The nonlinear collapse of Larmor precession depends critically on
the relative angle between the Larmor field and probe
polarization. Figure\ \ref{fig:angles} shows $1/e$ times for the initial decay, as a
function of $\theta $\ for an interval between $0^{\circ}$ and
$90^{\circ}$. The data supports each of the two qualitative aspects
noted above. First
the timescales for nonlinear collapse differ by a factor of two between $
\theta =0^{\circ}$ and $\theta =90^{\circ}$. Second, the $1/e$ decay time is
significantly increased at the critical angle $\theta \approx
54^{\circ}$, where the decay envelope is exponential rather than
Gaussian, signifying that the nonlinear term in the spin
Hamiltonian has been reduced to a level where it is no longer
visible in the dynamics. Also shown are the
predictions of our theoretical model, which are in good agreement
with our data over the full range of the angle $\theta $. It is
particularly interesting that the Faraday signal decay time at the
critical angle can be increased tenfold over the shortest
nonlinear collapse time and at least fourfold over the photon
scattering time, leading to a very significant increase in useful
measurement time. Figure \ref{fig:critical} shows further measured and calculated $1/e$
decay times vs. $\tau _{s}$\ at the critical angle. Both obey the
expected linear scaling over a wide interval, with the measured
decay times reaching a plateau just below 10 ms, most likely due
to dephasing caused by ensemble inhomogeneities. For scattering
times $\tau _{s}<1$ ms our data suggests that a near-optimal, i.e.\ decoherence limited measurement is possible.  

In summary we have performed a careful study of the interaction between an ensemble of laser cooled Cs atomic spins and a far detuned, linearly polarized probe beam, in the geometry used for Faraday measurements of the collective spin. The dynamics is dominated by a nonlinear term in the single-spin Hamiltonian arising from the probe light shift, which leads to rapid collapse and revivals in the expectation value for individual spins. The initial nonlinear collapse is much faster than the decay expected from probe photon scattering, and thus the most important factor limiting the useful time window available for a Faraday measurement. We have demonstrated that the nonlinearity can be cancelled in a measurement of Larmor precession, by orienting the probe polarization at a critical angle $\theta \approx 54^{\circ}$ with the magnetic field. This simple approach extended our useful measurement time by as much as a factor of ten, and provided insight that may prove helpful in other special cases. As an example, we are now exploring the use of polarization spectroscopy to probe the pseudospin associated with the $m=0$ clock doublet in Cs.  As shown by eq. 3, a bias field along the quantization axis renders the effective nonlinearity proportional to $\hat{F}_z^{2}$, which does not perturb the pseudospin or couple it to the rest of the ground manifold. To permit a true QND measurement of the spin-angular momentum when a suitable bias magnetic field cannot be applied, one can in principle use the light shift from a second, weak laser field to compensate for the nonlinearity induced by the probe.  If this laser is tuned between the two transitions of the $D_{1}$ line at 894 nm the resulting nonlinearity has the opposite sign of the far-detuned limit, Eq. 2. This, in combination with the large excited state hyperfine splitting, should allow the total nonlinearity to be cancelled without adding significantly to the overall rate of photon scattering. Alternatively, the stronger nonlinearity that can be achieved with a probe near the $D_{1}$ line can be used to drive and observe interesting coherent spin dynamics. We are currently pursuing this approach in an effort to realize a version of the kicked top \cite{twelve} based on an atomic spin, and hope to use this system to explore fundamental aspects of quantum chaos such as e.g.\ hypersensitivity to perturbation \cite{fourteen}. The general ability to design nonlinear dynamics will also allow one to extract information that goes beyond the mean value of the spin. This might then be used to implement new types of weak measurement, such as real-time estimation of the spin density matrix.

P. S. J. acknowledges the support of the National Science Foundation (PHY-0099582) and the Army Research Office (DAAD19-00-1-0375).  I. H. D. was supported by the National Science Foundation (PHY-0099569) and the Office of Naval Research (N00014-03-1-0508).

%\bibliography{continuus}

\end{document}